\documentclass[
aps,%
12pt,%
final,%
notitlepage,%
oneside,%
onecolumn,%
nobibnotes,%
nofootinbib,%
superscriptaddress,%
noshowpacs,%
centertags]%
{revtex4}

\begin{document}
\selectlanguage{russian}

\title{A Pulsar Time Scale Based on Parkes Observations in 1995--2010}

\author{\firstname{€.~….}~\surname{Rodin}}Rodin €.…. 
\email{rodin@prao.ru}
\affiliation{Pushchino Radio Astronomy Observatory, Astro Space Center, Lebedev Physical Institute, Russian Academy of Sciences, Pushchino, Moscow oblast, 142290 Russia}
\author{\firstname{V.~€.}~\surname{Fedorova}}Fedorova '.€.
\email{fedorova-astrofis@mail.ru}
\affiliation{Pushchino Radio Astronomy Observatory, Astro Space Center, Lebedev Physical Institute, Russian Academy of Sciences, Pushchino, Moscow oblast, 142290 Russia}
\affiliation{Pushchino State Natural Science Institute, Pushchino, Moscow oblast, Russia}


\begin{abstract}
{\bf Abstract} --- Timing of highly stable millisecond pulsars provides the possibility of independently verifying terrestrial time scales on intervals longer than a year. An ensemble pulsar time scale is constructed based on pulsar timing data obtained on the 64-m Parkes telescope (Australia) in 1995--2010. Optimal Wiener filters were applied to enhance the accuracy of the ensemble time scale. The run of the time--scale difference PT$_{ens}$ $-$ TT(BIPM2011) does not exceed $0.8 {\pm} 0.4$ $\mu s$ over the entire studied time interval. The fractional instability of the difference '$_{ens}$ $-$ TT(BIPM2011) over 15 years is ${\sigma_z} = (0.6\pm 1.6){\cdot}10^{-15}$, which corresponds to an upper limit for the energy density of the gravitational--wave background ${\Omega_g}h^2 {\sim} 10^{-10}$ and variations in the gravitational potential ${\sim} 10^{-15}$ Hz at the frequency $2{\cdot}10^{-9}$ ƒæ.
\end{abstract}

\maketitle

\section{INTRODUCTION}

Pulsar timing is a unique method that makes it possible to obtain information about completely different areas of science, from relativistic astrophysics and cosmology to fundamental metrology. We show in the current study that observations of millisecond pulsars can be used to construct an independent barycentric time scale whose stability is comparable to that of the best atomic standards over relatively long time intervals of 10--15 years.

Almost immediately after the discovery of pulsars, it became clear that the high regularity of the sequences of pulses emitted by pulsars can be used to construct a new astronomical time scale on this basis. It became obvious with the discovery of the first
millisecond pulsar \cite{backer} that this new astronomical time scale could compete with terrestrial atomic clocks in terms of its stability.

This pulsar time scale is long-term, unified, reproducible, and indestructible for all observers on Earth. The long lifetimes of pulsars ($10^8$ yrs) make it possible to compare realizations of terrestrial atomic time scales for different years. Since pulsars are located outside the solar system, they provide the only means of independently verifying the terrestrial atomic time scale, which is impossible to do purely by intercomparing terrestrial clocks. We see precisely this to be the main role of a pulsar time scale.

According to intergovernmental standard (GOST) 8.567-99, the pulsar time scale is defined to be a continuous sequence of time intervals between pulses measured from an initial time $t_0$ and expressed by the formula

\begin{equation}
t_N = t_0 +P_0 N + \frac{1}{2} P_0 \dot{P} N^2 +u(t) + v(t),
\end{equation}
where $P_0$ and $\dot{P}$ are the period and period derivative at time $t_0$, and $u(t)$ and $v(t$) are additive noise due to variations in the pulsar rotation and variations in the running of a reference atomic frequency standard, respectively. The residual deviations in the pulse arrival times (PATs) $r(t) = u(t) + v(t)$ remaining after subtraction of the model from the observations contain information about physical processes occurring in the pulsar and the frequency standard. It is supposed based on general physical considerations that the random processes $u(t)$ and $v(t)$ are uncorrelated. We also assume that the noise runs $u(t)$ for different pulsars are likewise uncorrelated. The construction of an ensemble pulsar time scale reduces to estimating the signal $v(t)$ in the presence of the additive interference $u(t)$.

As an example of a cosmological application of pulsar timing, we should mention first and foremost the 1978 paper of Sazhin \cite{sagh}, where it was proposed to use an analysis of PATs to search for gravitational waves. This work continues to be cited in studies dedicated to pulsar timing.

Detweiler \cite{det} published a study on a similar topic in 1979, but emphasizing searches for stochastic­  gravitational-wave radiation. The first estimated upper limit to the energy density of the stochastic gravitational-wave background relative to the critical density of the Universe was $\Omega_g h^2 \sim 1$ at a frequency of $3 \cdot 10^{-8}$ Hz. 

Hellings and Downs \cite{hd} also gave an upper limit for $\Omega_g h^2$ at a frequency $< 10^{-8}$ Hz at the level $1.4 \cdot 10^{-4}$.

\section{OBSERVATIONS}

The pulsar observations we have used to construct an ensemble pulsar time scale were obtained on the 64-m Parkes radio telescope (Australia) at 0.7, 1.4, and 3.1 GHz during 1995--2010 as part of the Parkes Pulsar Timing Array (PPTA) program \cite{man}. Files with data suitable for directly determining the residual PAT deviations, as well as the rotational and astrometric parameters of the pulsars, are available at the web site data.csiro.au\footnote{https://data.csiro.au/dap/ landingpage?execution=e3s2}. We used the Tempo2 software \cite{Hobbs} to reduce the topocentric PATs to the solar system
barycenter. No fitting of the pulsar parameters was conducted.

\section{DATA REDUCTION}

The algorithm used to reduce the topocentric PAT values to the solar system barycenter is based on the difference between the topocentric $\tilde{t}$ and solar system barycentric $t_N$, arrival times of a given pulsar pulse, which can be written \cite{mar}

\begin{equation}
c(t_N - \tilde{t}) = \vec{k}\cdot \vec{r} - \frac{1}{2R} [\vec{k} \times \vec{r}]^2 + \gamma + \frac{DM}{f^2},
\end{equation}
where the first term on the right-hand side of (2) -- the scalar product of the barycentric unit vector $\vec{k}$ and the radius vector of the observer $\vec{r}$ -- makes the main contribution and is called the Roemer correction; the second, third, and fourth terms are corrections for the sphericity of the front, the relativistic delay of the signal in the gravitational field of the solar system, and
the delay of the signal in the interstellar and interplanetary plasma; and $DM$ is the dispersion measure in the direction toward the pulsar, $R$ the distance to the pulsar, $f$ the observing frequency, and $c$ the speed of light.

The barycentric arrival time $t_N$ of the $N$th pulse is calculated using the rotational parameters of the pulsar $P_0$ and $\dot{P}$ with formula (1). The planetary
epherides DE421 were used to reduce the topocentric PATs to the barycenter.

The standard algorithm for the formation of the ensemble pulsar time scale involves measuring the PATs from several pulsars relative to the same reference scale and calculating the weighted sum $\sum_{i} w_i r_i \approx{v(t)}$, where $w_i$ is the relative weight ascribed to pulsar $i$. It is assumed that the rotational variations $u_i$ of each of the pulsars are also uncorrelated, and will tend to mutually cancel when averaged.

An algorithm for the formation of an ensemble pulsar time scale based on Wiener filtration is proposed in \cite{rodin,RODIN,rch}, where it is shown that applying optimal filters makes it possible to separate the contributions to $r$ of the noises $u_i$ and $v_i$ and thereby form an ensemble pulsar time scale at a higher level of accuracy.

The following method can be applied to separately find the variations of the pulsar rotations and of the reference time scale of the frequency domain. the Fourier transform of the residual deviations for the $i$th ($i = 1,2, ..., M$) are first calculated:

\begin{equation}
x_i (w) = \frac{1}{\sqrt{n}} \sum_{t = 1}^{n} r_i e^{i(w-1)(t-1)},
\end{equation}

The power spectra ($i = j$) and cross spectra ($i \neq j$) are then calculated using the formula

\begin{equation}
X_{ij} = \frac{1}{2 \pi}|x_i(w)x_j(w)|,
\end{equation}

We have $\frac{M(M-1)}{2}$ cross spectra for $M$ pulsars. The desired signal $v_i(t)$ can be calculated as follows:

\begin{equation}
v(t) = \mathcal{F}^{-1} \Bigg[{\sqrt{\frac{\bar{S_v}}{S_{ui} + S_{vi}}}}T_i(f)\Bigg],
\end{equation}
where $S_{ui}$ and $S_{vi}$ are the power spectra of the random processes $u_i(t)$ and $v_i(t)$ and $T_i(f)$ is the Fourier transform of $r_i(t)$, $\mathcal{F}^{-1}$ is the inverse Fourier transform operator. The quantity $\bar{S_v}$ is the averaged cross spectrum

\begin{equation}
\bar{S_v}(u) = \frac{2}{M(M-1)} \sum_{m = 1}^{\frac{M(M-1)}{2}} X_m(w),
\end{equation}
which serves as an estimate of the power spectrum of the signal $v(t)$.The ensemble time scale $v_{ens}(t)$ can be written

\begin{equation}
v_{ens} (t) = \sum_{i=1}^{M} w_i v_i(t)
\end{equation}

The calculation of the relative weights of the pulsars $w_i$ requires a separate discussion, since the behavior and stability of the ensemble time scale depends strongly on the choice of weights. We chose two options for calculating the weights $w_i$:
 
\begin{enumerate}
\item $w_i = const$;
\item $w_i = \varkappa \sigma_{ri}^{-2}$, where the normalization coefficient $\varkappa$ serves to ensure satisfaction of the condition $\sum_{i} w_i = 1 (i = 1, 2, ..., M)$, and $\sigma$ is the rms deviation of the series.
\end{enumerate}

\section{RESULTS}
As a first step in constructing an ensemble pulsar time scale using 20 pulsars observed in the PPTA program, we first visually selected eight pulsars having the longest series of observations and the minimum noise levels in the residual deviations: J0437--4715, J1024--0719, J1603--7202,J1713+0747, J1730--2304, J1744--1134, J2124--3358, and J2145--0750. After calculating the rms deviations, the series of residual deviations for J2124--3358 were excluded, since it displayed an enhanced level of rms deviations.

All the data were averaged over 30 days. Occasional gaps in the measurements were filled via a linear interpolation of the neighboring values. A common observational interval was used for all the pulsar series. The length of the series after reducing
them to a homogeneous form was 181 points. Since the length of the series had changed, we again fitted them with a quadratic polynomial. Figure 1 shows the barycentric residual PAT deviations $v_i(t)$ for the seven pulsars after they have passed through an optimal Wiener filter. The bold solid and dashed curves show the weighted averaged and simple averaged ensemble time scales, respectively. We applied a low--frequency
filter to the ensemble time scales to smooth high--frequency variations. The transmission bandwidth of the filter was determined visually, in order to distinguish relatively long-term variations with periods $P > 0.5$ yrs. The variations of the overall run do not exceed 0.8 $\mu $s. We can also see that the accuracy of the timing improves with time: the mean uncertainty in the calculated ensemble time scale was reduced from 0.4 to 0.3 $\mu s$ after 2002.

A plot of the fractional instability of the frequency standard$\sigma_z$ an be used to estimate the deviations of its frequency from the nominal value over various time intervals \cite{taylor}. Figure 2 (upper) shows the behavior of the fractional instability of the seven millisecond pulsars as a function of the averaging interval $\tau$. The lower panel of Fig. 2 shows the run of $\sigma_z$, calculated $1$ with the weights $w_i = \varkappa \sigma_{ri}^{-2}$. (dashed curve) and 2 with equal weights (solid bold curve), together with 3 the run of $\sigma_z$ from Hobbs et al. \cite{Hobbs} (thin solid curve).

\section{DISCUSSION}
We have constructed an ensemble pulsar time scale PT$_{ens}$. based on pulsar timing data obtained at Parkes (Australia) in 1995--2010. The fractional instability of the difference PT$_{ens}$ - TT(BIPM2011) over a 15--year interval is $\sigma_z \approx (0.6 \pm 1.6) \cdot 10^{-15}$. Further refinement and enhancement of the accuracy of pulsar timing would make it possible to decrease this value by a factor of three to five, achieving the level $\sigma_z \sim (1 \div 2) \cdot 10 ^ {-16}$. In our opinion, further progress in improving the stability of pulsar time scales will require appreciable effort, due to the limitations imposed by the physical nature of the pulsar rotations.

Applying different methods for weighting the series when constructing the ensemble time scale leads to appreciable differences in its behavior over various time intervals. For example, the version of $\sigma_z$ calculated via a simple averaging of the filtered data displays better stability in the interva $\tau < 11$yrs (curve 2 in the lower panel of Fig. 2) than the version of$\sigma_z$calculated using the weights $w \sim \sigma_{r}^{-2}$ (curve $1$). This result has important practical value. For example, if
an ensemble pulsar time scale is used as a reference scale for space missions with durations of less than a few years, it is possible to form this scale using a simple averaging, without being concerned about the relative weighting of individual pulsars.

A detailed analysis of the weights $w_i, (i = 1, 2, ..., M)$ showed that the pulsar J0437-4715 has the relative weight $w_i = 0.62$, so that the weighted ensembletime scale $"$hangs$"$ to an appreciable extent on this one pulsar, which we believe to be a shortcoming of this method for weighting the data.

The variations of the run of the ensemble pulsar time scale PT$_{ens}$ can have the following interpretations.
\begin{enumerate}
\item Variations of the run of the reference terrestrial time scale TT are due to instability of the run of the primary frequency standards used to form the TT scale.
\item Variations of the gravitational potential in the vicinity of the solar system lead to variations in the rates of all processes on the Earth, including those occurring in atomic frequency standards.
\end{enumerate}

In connection with the latter interpretation, we note the work of Khmelnitsky and Rubakov \cite{Khmelnitsky}, who related variations of the gravitational potential with
oscillations in the pressure of a super--light scalar field, whose particles, bosons, do not interact with either matter or each other. The pressure of this scalar field is non-zero, and oscillates about zero, which can be interpreted physically as oscillations of the potential with a frequency directly proportional to the mass of the boson. They give an estimate of the potential oscillations at a level of $10^{-15}$ at nanohertz frequencies. The signal from dark matter in the form of a scalar field possesses two distinguishing properties: first, this signal does not depend on the direction toward the pulsar and second, it is monochromatic, with a frequency that depends on the mass of the dark--matter particles. These properties may enable the detection of this signal through an analysis of the ensemble pulsar time scale.

Porayko and Postnov \cite{Post} computed an upper limit for the amplitude of the variable gravitational potential in a monochromatic approximation based on pulsar timing data from the NANOGrav project \cite{Ferdman}, which they found to be $\Psi_c < 1.14\cdot 10^{-15}$ at a frequency of $f = 1.75 \cdot 10^{-8}$ Hz.

In relation to estimates of variations of the gravitational potential, we especially note the properties of calculations of the fractional instability $\sigma_z(\tau)$. This
quantity displays the amplitude of a cubic polynomial fitted to the data taken on increasing intervals  $\tau$ \cite{taylor}. In other words, $\sigma_z$ gives an estimate of the integrated effect of the run of one scale relative to another (for
example, TT relative to PT$_{ens}$ in an interval $\tau$) due to various physical effects, including variations of the gravitational potential in the vicinity of the solar system. Thus, $\sigma_z$ can be used directly to measure such effects.

\section{CONCLUSION}
We have constructed an ensemble pulsar time scale based on timing data for millisecond pulsars obtained on the Parkes telescope (Australia) in 1995--2010. Analysis of the relative instability of the run of the ensemble pulsar time scale shows that the quantity $\sigma_z$ is at the level $(0.6\pm 1.6){\cdot}10^{-15}$ over this 15--year interval, corresponding to an upper limit for the energy density of the relict gravitational--wave background $\Omega_g h^2 \sim 10^{-10}$ at a frequency of $2 \cdot 10^{-9}$ Hz, and oscillations of the gravitational potential $\sim 10^{-15}$ at this same frequency. Analysis of the run of the fractional instability $\sigma_z(\tau)$ shows
the acceptability of using simple averages of the individual scales based on the rotations of individual pulsars to contruct the ensemble pulsar time scale on relatively short intervals $\tau \lesssim 10 $ yrs. A weighted average must be used on longer intervals $\tau \gtrsim 10 \div 15$ yrs.

This work was supported by the program of the Presidium of the Russian Academy of Sciences $"$Transitional and Explosive Processes in Astrophysics$"$ and the Russian Foundation for Basic Research (grant 16-02-00954).

Translated by D.Gabuzda.

\appendix

\newpage
\begin{figure}[h!]
\setcaptionmargin{1mm}
\vbox{\includegraphics[width=1\linewidth]{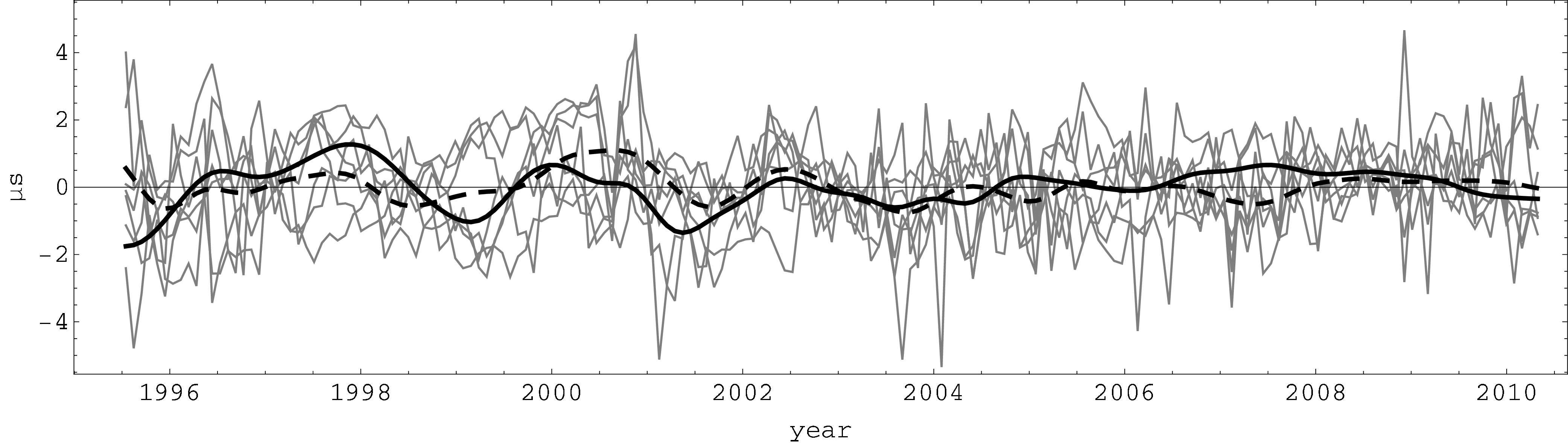}}
\caption{Barycentric residual PAT deviations of seven millisecond pulsars relative to the time scale TT(BIPM2011) passed  through an optimal Wiener filter. The bold solid curve shows the ensemble pulsar time scale calculated with weights $w_i \sim \sigma_{ri}^{-2}$, based on the filtered data. The uncertainty in the ensemble time scale is 0.4 $\mu s$. The bold dashed curve shows the time scale obtained via a simple average of the filtered data.}
\label{ris:fig1}
\end{figure}

\newpage
\begin{figure}[h!]
\begin{minipage}[h]{0.49\linewidth}
\center{\includegraphics[width=1.1\linewidth]{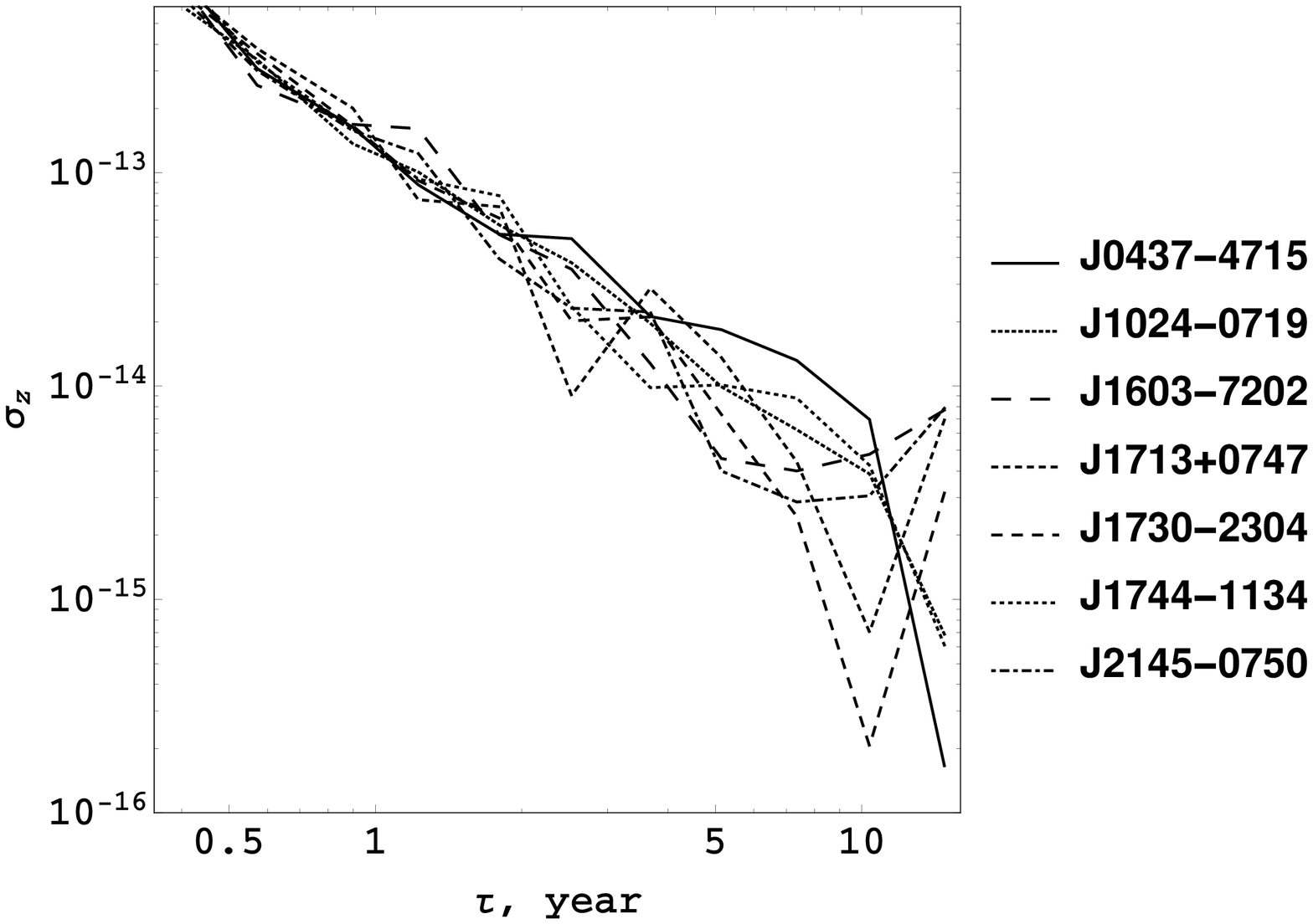} \\ a)}
\end{minipage}
\hfill
\setcaptionmargin{1mm}
\begin{minipage}[h]{0.49\linewidth}
\center{\includegraphics[width=0.9\linewidth]{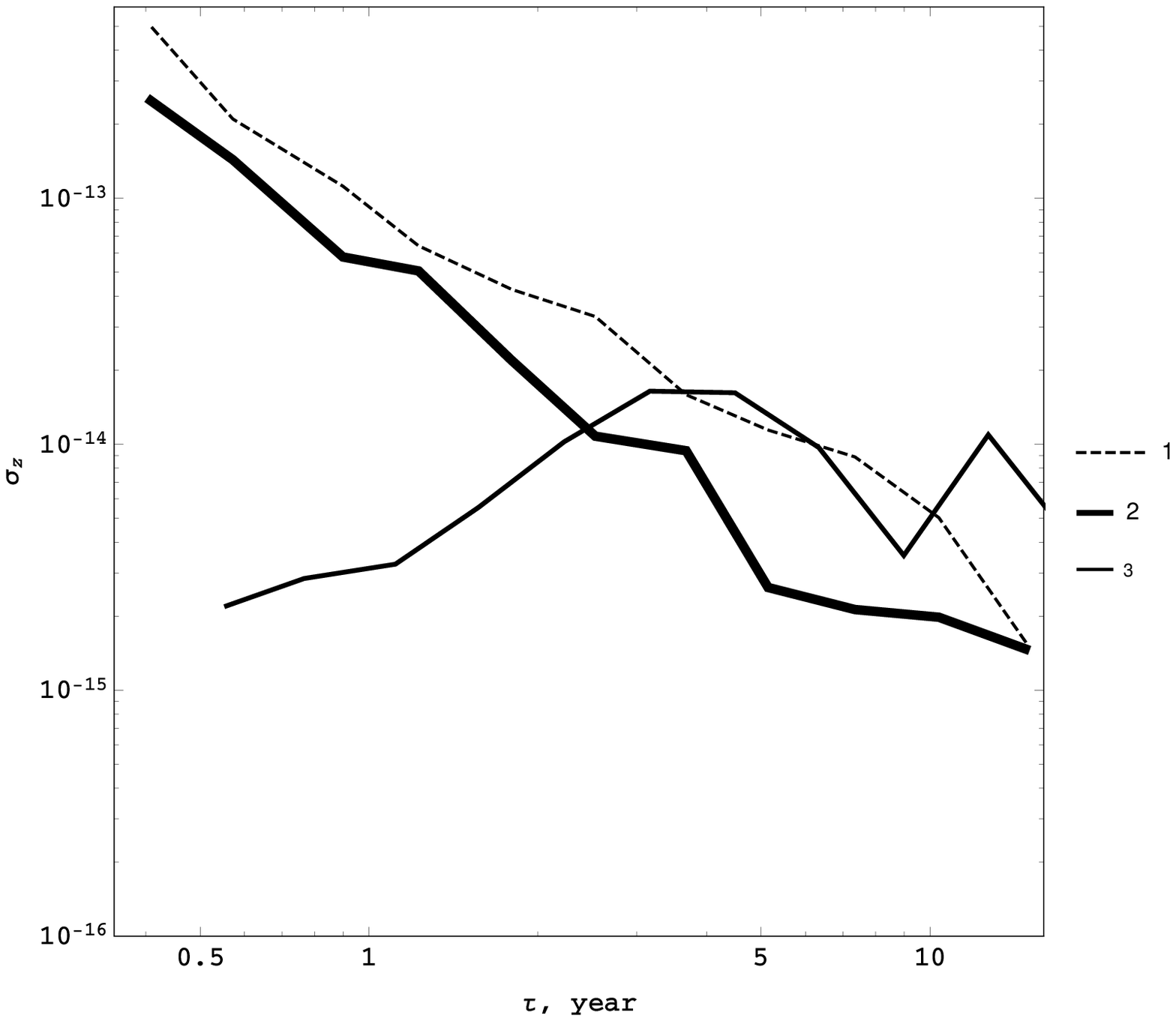} \\ b)}
\end{minipage}
\label{ris:image1}
\caption{Plots of the fractional instability $\sigma_z$ of the seven millisecond pulsars (upper) and of the ensemble pulsar time scale (lower), constructed in various ways: with weights $w_i = \varkappa \sigma_{ri}^{-2}$  $\it (1)$, with equal weights $\it (2)$, and in accordance with the computations of \cite{Hobbs} $\it (3)$. The fractional instability of the ensemble time sale constructed using the filtered, weighted data over the 15-year time interval is $\sigma_z = (0.6 \pm 1.6) \cdot 10^{-15}$.}
\label{ris:fig2}
\end{figure}

\end{document}